\documentclass{article}
\usepackage[utf8]{inputenc}
\usepackage{amsmath,amssymb,bbm}
\usepackage{preamble}

\title{Differential Privacy in the Shuffle Model:\\ A Survey of Separations}
\author{Albert Cheu}

\begin{document}

\maketitle

\section{Introduction}

Many differentially private algorithms operate in the \emph{central model}, also known as the trusted curator model. Here, a single analyzer has raw user data and its computations are insensitive to any one user's data point. But the fact that all users give their data to one party means that there is a single point of failure: the privacy of all users is contingent on the integrity of the analyzer.

There are a number of ways to model weaker trust in the analyzer. Perhaps the most well-known among them is the \emph{local model}. Here, the dataset is a distributed object where each user holds a single element. To preserve their own privacy, each user randomizes their data point and submits the output to the analyzer. Because the signal from each user is hidden behind noise, there are a number of lower bounds on the error of locally private protocols that strongly separate the local model from the central model \cite{BNO08,CSS12,BS15,ACFT19,JMR19}. That is, the analyzer needs more samples (users) to achieve the same accuracy as in the central model. Locally private protocols are also more vulnerable to manipulation: by sending carefully distributed messages, malicious users can skew tests and estimates of distributions beyond simply changing the input of the protocol \cite{CSU19}. These negative results lead us to ask the following question:
\begin{quotation}
Can we achieve the accuracy that is possible with centrally private algorithms

from a trust assumption that is close to locally private protocols?
\end{quotation}

Research into the \emph{shuffle model} has given an answer to this question. Like the local model, users in a shuffle protocol produce messages by feeding their data into a local randomizer. But now they trust some entity to apply a uniformly random permutation on all user messages. We assume that the adversary's view is limited to that permutation, so no message can be linked back to its sender.

This survey gives an overview of the recent surge of work in the shuffle model. We pay particular attention to results that characterize the strength of the model relative to the local and central models.

\paragraph{Outline.} In Section 2, we establish the requisite privacy and model definitions. In Section 3, we contrast local model lower bounds with shuffle model upper bounds: there are problems for which additive error and sample complexity are much lower in the shuffle model. Then, in Section 4, we give techniques to show that the shuffle model (under natural constraints) is weaker than the central model. Finally, Section 5 gives a glimpse of what is possible in interactive variants of the model.

All these results focus on the accuracy of differentially private shuffle protocols. In Appendix \ref{apdx:other-models}, we explore alternative models and compare them with the shuffle model. Appendix \ref{apdx:comm} contains an overview of shuffle protocols that are designed with the aim of reducing costs of transmission (e.g. number of messages and total number of bits consumed by messages). And Appendix \ref{apdx:brittle} highlights two unusual shuffle protocols which pose a challenge to proving lower bounds.

\paragraph{Author's Note.} Much of this survey is derived from the author's PhD. thesis. Notation and definition changes have been introduced to simplify the presentation

\section{Preliminaries}

We will use the notation $[d] = \{1, 2, \ldots, d\}$, $\mathbb{N} = \{1, 2, \ldots \}$. A dataset $\vec{x} \in \cX^n$ is an ordered tuple of $n$ rows where each row is drawn from a data universe $\cX$ and corresponds to the data of one user. Two datasets $\vec{x},\vec{x}\,' \in \cX^n$ are considered \emph{neighbors} if they differ in at most one row. This is denoted as $\vec{x} \sim \vec{x}\,'$.

\begin{definition}[Differential Privacy \cite{DMNS06}]
An algorithm $\cM: \cX^n \rightarrow \cZ$ satisfies \emph{$(\eps, \delta)$-differential privacy} if, for every pair of neighboring datasets $\vec{x}$ and $\vec{x'}$ and every subset $T \subset \cZ$,
	$$\pr{}{\cM(\vec{x}) \in T} \le e^\eps \cdot \pr{}{\cM(\vec{x}\,') \in T} + \delta.$$
	
When $\delta > 0$, we say $\cM$ satisfies \emph{approximate} differential privacy. When $\delta = 0$, $\cM$ satisfies \emph{pure} differential privacy and we omit the $\delta$ parameter.
\end{definition}

Because this definition assumes that the algorithm $\cM$ has ``central'' access to compute on the entire raw dataset, we sometimes call this \emph{central} differential privacy.

The \emph{binomial mechanism} is a centrally private algorithm that has proven to be useful in the design and analysis of shuffle protocols.

\begin{lemma}[Binomial Mechanism \cite{DKMMN06,GGK+19}]
\label{lem:b_noise}
	Let $f \colon \cX^n \to \Z$ be a 1-sensitive function, i.e. $|f(\vec{x}) - f(\vec{x}\,')| \leq 1$ for all neighboring datasets $\vec{x}, \vec{x}\,' \in \cX^n$. There is a constant $\kappa$ such that, for any $\ell\in\N$, $p\in(0,1)$, and $\eps, \delta \in (0,1)$ satisfying $$\ell \cdot \min(p,1-p) \geq \frac{\kappa}{\eps^2} \cdot  \log \frac{1}{\delta},$$ the algorithm that samples $\eta \sim \Bin(\ell, p)$ and outputs $f(\vec{x}) + \eta$ is $(\eps,\delta)$-differentially private. The error is $O\left(\tfrac{1}{\eps}\sqrt{\log \tfrac{1}{\delta}}\right)$ with constant probability.
\end{lemma}



\subsection{The Local Model}
We first establish the local model. Here, the dataset is a distributed object where each of $n$ users holds a single row. Each user $i$ provides their data point as input to a randomizing function $\cR$ and publishes the outputs for some analyzer to compute on.

\begin{definition}[Local Model \cite{Warner65,EGS03}]
A protocol $\cP$ in the \emph{local model} consists of two randomized algorithms:
\begin{itemize}
    \item A randomizer $\cR:\cX \times \zo^r \to\cY$ mapping a data point and public random bits to a message
    \item An analyzer $\cA:\cY^n \times \zo^r \to\cZ$ that computes on a vector of messages and public random bits
\end{itemize}
We define its execution on input $\vec{x}\in\cX^n$ as $$\cP(\vec{x}) := \cA(\cR(x_1,W),\dots,\cR(x_n,W)),$$ where $W$ is a uniformly random member of $\zo^r$.
\end{definition}

It is possible to extend the model definition to allow for multiple rounds of communication. To ease readability, we defer discussion of interactive protocols (local and shuffle) to a later section.

\medskip

Suppose the privacy adversary wishes to target user $i$. In this model, the adversary's view is limited to the output of $\cR(x_i,W)$ so we impose the privacy constraint on $\cR$.

\begin{definition}[DP in the Local Model \cite{DMNS06,KLNRS08}]
A protocol $\cP=(\cR,\cA)$ is $(\eps,\delta)$-\emph{local differentially private} if, for all $w \in \zo^r$, $\cR(\cdot, w)$ is $(\eps,\delta)$-differentially private. That is, the privacy guarantee is over the internal randomness of the users' randomizers and not the public randomness of the protocol.
\end{definition}




\subsection{The Shuffle Model}
To give intuition for the shuffle model, we start by sketching a preliminary version called the \emph{single-message shuffle model}. Like the (one-round) local model, users execute $\cR$ on their data to produce messages but users now trust a service to perform a secure shuffle on the messages. That is, an adversary's view is limited to a uniformly random permutation of the messages, so no message can be linked back to its sender. Intuitively, whatever privacy guarantee is granted by $\cR$ is \emph{amplified} by this anonymity: to learn about $x_i$, an adversary has to not only recover information from one noisy message $y_i$ but somehow identify the target message inside a vector $\vec{y}$ of $n$ messages. \emph{Amplification-by-shuffling} lemmas quantify how well the privacy parameters are improved \cite{EFMRTT19,BBGN19,FMT20}. These lemmas provide a simple way to design protocols in the single-message shuffle model.

But the amplification lemmas do not apply to the relaxed version of the model where each user sends any (possibly randomized) number of messages to the shuffler. Here, we assume the shuffling prevents messages from the same sender from being linked with one another. We give a formal definition below:

\begin{definition}[Shuffle Model \cite{BittauEMMRLRKTS17, CSU+19}]
A protocol $\cP$ in the \emph{shuffle model} consists of three randomized algorithms:
\begin{itemize}
\item
    A \emph{randomizer} $\cR: \cX \times \zo^r \rightarrow \cY^*$ mapping a data point and public random bits to (possibly variable-length) vectors. The length of the vector is the number of messages sent. If, on any input, the probability of sending $m$ messages is 1, then we have an \emph{$m$-message protocol}.
\item
    A \emph{shuffler} $\cS: \cY^* \rightarrow \cY^*$ that concatenates message vectors and then applies a uniformly random permutation to the messages. 
\item
    An \emph{analyzer} $\cA: \cY^*\times \zo^r \rightarrow \cZ$ that computes on a permutation of messages and public random bits.
\end{itemize}
As $\cS$ is the same in every protocol, we identify each shuffle protocol by $\cP = (\cR, \cA)$. We define its execution on input $\vec{x}\in\cX^n$ as
$$
\cP(\vec{x}) := \cA(\cS(R(x_1,W), \ldots, R(x_n,W))),
$$
where $W$ is again the public random string. We assume that $\cR$ and $\cA$ have access to $n$.
\end{definition}

\begin{rem}
By making $n$ accessible to the parties, we allow internal parameters to depend on $n$. This enables users to evenly distribute the responsibility of adding noise.
\end{rem}

With this setup, we use the following definition of shuffle differential privacy.
\begin{definition} [DP in the Shuffle Model \cite{CSU+19}]
\label{def:shuffle_dp}
	A shuffle protocol $\cP = (\cR, \cA)$ is \emph{$(\eps, \delta)$-differentially private} if, for all $w\in\zo^r$ and all\footnote{Some protocols assume lower bounds on $n$ in order to invoke concentration arguments. These bounds will typically be small.} $n\in \N$, the algorithm $$(\cS \circ \cR^n)(\vec{x}) := \cS(\cR(x_1,w), \ldots, \cR(x_n,w))$$ is $(\eps, \delta)$-differentially private.
\end{definition}

For brevity, we typically call these protocols ``shuffle private.'' We will also drop the public randomness input if it is unused.

Note that Definition~\ref{def:shuffle_dp} assumes all users follow the protocol. Ideally, distributed protocols should still guarantee some level of privacy even when users are malicious. A simple attack is to drop out: let $\cS\circ\cR^{\gamma \cdot n}$ denote the case where a $\gamma$ fraction of $n$ users execute $\cR$ but they are only given access to $n$ (not $\gamma$). $\cS\circ\cR^{1\cdot n}$ might satisfy a particular level of differential privacy but there could be a value $\gamma < 1$ where $\cS\circ\cR^{\gamma\cdot n}$ does not.\footnote{Note that, with respect to differential privacy, dropping out is the worst malicious users can do. This is because adding messages from malicious users to those from honest users is a post-processing of $\cS\circ \cR^{\gamma n}$. If $\cS\circ \cR^{\gamma n}$ is already differentially private for the outputs of the $\gamma n$ users alone, then differential privacy's resilience to post-processing ensures that adding other messages does not affect this guarantee. Hence, it is without loss of generality to focus on drop-out attacks.} This motivates a definition of shuffle privacy that is \emph{robust} to a malicious minority of users:

\begin{definition} [Robust DP in the Shuffle Model]
\label{def:robust_shuffle_dp}
    A shuffle protocol is $(\eps,\delta)$-robustly differentially private if, for all $w\in\zo^r$, the algorithm $$(\cS \circ \cR^{n/2})(\vec{x}) := \cS(\cR(x_1,w), \ldots, \cR(x_{n/2},w))$$ is $(\eps, \delta)$-differentially private.
\end{definition}

\paragraph{Discussion of Definition.} We have defined robustness with regard to privacy rather than accuracy. A robustly shuffle private protocol promises its users that their privacy will not suffer much from a limited fraction of malicious users. But it does not make any guarantees about the accuracy of the protocol; we will state our accuracy guarantees under the assumption that all users follow the protocol.

Also, observe that robustness is not immediately implied by the basic form of shuffle privacy in Definition~\ref{def:shuffle_dp}. Appendix \ref{apdx:brittle} describes protocols that satisfy shuffle privacy but are not robust to drop-outs.

Finally, the constant $1/2$ in the definition (corresponding to the assumption of an honest majority) can be changed to an arbitrary constant without changing the asymptotics of the upper or lower bounds.

\paragraph{Comparison with prior definitions.} We remark that the work by Balcer, Cheu, Joseph, and Mao \cite{BCJM20}---which originally formalized robustness in the shuffle model---offers a definition where privacy parameters are functions of the unknown fraction of users who are honest. The functions should be continuous and non-increasing, meaning that the privacy guarantee gradually loosens from $(\eps,\delta)$ to $(O(\eps),O(\delta))$. Although more general, their definition demands more notation which we avoid for simplicity.

We also note that Definition \ref{def:robust_shuffle_dp} is essentially an adaptation of earlier work by \'{A}cs and Castelluccia \cite{AC11} on distributed differential privacy.

\section{Separations between Local \& Shuffle Privacy}
In this section, we will introduce four problems. For each problem, we will state a lower bound in the local model and then describe a protocol in the shuffle model that breaks through that bound. To simplify the presentation, we will assume $\eps<1$ and $\delta = O(1/\poly(n))$.

\begin{table}[h]
\renewcommand*{\arraystretch}{1.8}
    
    \centering
    \begin{tabular}{ccccc}
         & Local & Shuffle \\ \hline
        Error of & $\Omega\paren{\frac{1}{\eps} \sqrt{n}}$ & $O\paren{\frac{1}{\eps} }$\\
        Binary Sums & \cite{BNO08,CSS12} & \cite{BBGN19-2,CY21} \\ \hline
        $\ell_\infty$-error of & $\Omega\paren{\frac{1}{\eps} \sqrt{n \log k}}$ & $O\paren{\frac{1}{\eps^2} \log \frac{1}{\delta} }$ \\
        $d$-bin Histograms & \cite{BS15} & \cite{BC19} \\ \hline
        Sample Complexity of & $\Omega\paren{\frac{d}{\alpha^2\eps^2}}$ & $O\paren{\frac{d^{2/3}}{\alpha^{4/3} \eps^{2/3}} + \frac{d^{1/2}}{\alpha \eps} + \frac{d^{1/2}}{\alpha^2}}$ \\
        $\alpha$-Uniformity Testing & \cite{ACFT19} & \cite{CY21} \\ \hline
        Sample Complexity of & $\Omega(\ell)$ & $O\paren{\frac{1}{\eps^2}\log \frac{1}{\delta}}$ \\
        $(2,\ell)$-Pointer-Chasing & \cite{JMR19} & \cite{BC19} \\ \hline
    \end{tabular}
    
    \caption{Lower bounds in the local model aligned with upper bounds in the shuffle model}
    \label{tab:upper}
\end{table}

\subsection{Binary Sums}
In this problem, each user $i$ has a bit $x_i\in \zo$ and the objective is to compute the sum. Dating back to Warner \cite{Warner65}, \emph{randomized response} is the canonical local protocol for this problem. The randomizer is below:
\[
\cR_\rr(x_i) := \begin{cases}
\Ber(1/2) & \textrm{with probability } p \\
x_i & \textrm{otherwise}
\end{cases}
\]
Let $y_i$ be the message sent by user $i$. Due to subsampling and noise addition, the expected value of $\sum y_i$ is $(1-p) \cdot \sum x_i + np/2$. The analyzer will re-center and re-scale to obtain an unbiased estimator:
\begin{align*}
\cA_\rr(\vec{y}) &:= \frac{1}{1-p} \left( \sum y_i -np/2 \right)\\
\ex{}{\cA_\rr(\vec{y})} &= \frac{1}{1-p} \cdot \left( \ex{}{ \sum y_i } - np/2 \right) \\
    &= \sum x_i
\end{align*}
Setting $p\gets 2/(e^\eps+1)$ suffices for $\eps$-local privacy but incurs an additive error of $O(\frac{1}{\eps} \sqrt{n})$. This is optimal.
\begin{thm}[Beimel et al. \cite{BNO08} \& Chan et al. \cite{CSS12}]
Let $\cP$ be an $(\eps,\delta)$-locally private protocol. If $\cP$ computes binary sums up to additive error $\alpha$ with constant probability, then $\alpha = \Omega(\frac{1}{\eps} \sqrt{n})$.
\end{thm}

\subsubsection{Shuffling Randomized Response}

Note that $\cP_\rr := (\cR_\rr,\cA_\rr)$ can also be interpreted as a single-message shuffle protocol. Cheu et al. \cite{CSU+19} show that the parameter $p$ can be chosen such that $\rr$ achieves robust shuffle privacy while also avoiding error that scales polynomially with $n$.
\begin{thm}[Cheu et al. \cite{CSU+19}]
There exists a choice of $p$ such that the shuffle protocol $\cP_\rr = (\cR_\rr, \cA_\rr)$ is $(\eps, \delta)$-robustly private and computes binary sums up to additive error $O(\frac{1}{\eps} \sqrt{\log \frac{1}{\delta}})$ with constant probability.
\end{thm}

\begin{proof}[Proof]
We will set $p$ to a value $\Omega(\frac{1}{\eps^2 n} \log \frac{1}{\delta})$. If this quantity exceeds $1/2$ (which occurs when $n$ is not large enough), $p$ must take a different form and the analysis will naturally change; we omit this technicality for neatness. Refer to \cite{CSU+19} for more details.

\underline{Robust privacy}: Assume without loss of generality that the honest majority is the set $[n/2]$. We leverage the fact that the view of an adversary is an unordered set of bits. This object contains as much information as the sum of those bits. More formally, given $\sum_{i=1}^{n/2} y_i$, the adversary can simulate a sample from $\cS(y_1,\dots,y_{n/2})$: pick a uniformly random binary string of length $n/2$ and sum $\sum_{i=1}^{n/2} y_i$. This procedure is a post-processing operation, which means we only have to ensure the privacy of $\sum_{i=1}^{n/2} y_i$.

By construction, some set of users $H\subset [n/2]$ will report messages sampled from $\Ber(1/2)$ and the rest will report their true values. So for any fixed set $H$, $\sum_{i=1}^{n/2} y_i$ is a sample from $\sum_{i \in [ n/2] - H} x_i + \Bin(|H|,1/2)$. If we show $|H| \geq \frac{\kappa }{\eps^2} \cdot  \log \frac{1}{\delta}$ where $\kappa$ is the constant in Lemma \ref{lem:b_noise}, then we can invoke the binomial mechanism to conclude that $\cS\circ\cR^{n/2}$ satisfies $(\eps, \delta)$-differential privacy.

Membership in $H$ is a Bernoulli process, so $|H|\sim \Bin( n/2, p)$. Due to our choice of $p$, standard concentration arguments imply $|H| \geq \frac{\kappa }{\eps^2} \cdot \log \frac{1}{\delta}$ with at least $1-\delta$ probability.

\underline{Accuracy}: We bound the protocol's error under the assumption that all users are honest. Recall that the output of the protocol is $\frac{1}{1-p} (\sum y_i -np/2)$. By a Chernoff bound, we have that $\sum y_i -np/2 $ is within $O(\frac{1}{\eps} \sqrt{\log \frac{1}{\delta}} )$ of its expectation. And because $\frac{1}{1-p} < 2$, the error of the unbiased estimator is $O(\frac{1}{\eps} \sqrt{\log \frac{1}{\delta}} )$.
\end{proof}

\paragraph{The Noise/Data Dichotomy.} In randomized response, a message can either be a Bernoulli bit or a data bit. Balle, Bell, Gasc{\'{o}}n, and Nissim \cite{BBGN19} prove that this dichotomy is in fact one instance of a general phenomenon: \emph{any} locally private randomizer can be expressed as a mixture of noise and data. Specifically, there is a ``blanket'' distribution $\bB$ and a parameter $p$ such that, for any input $x$, the distribution of $\cR(x)$ is equal to $p\bB + (1-p)\bD_x$ where $\bD_x$ an input-dependent distribution (identity in the case of $\cR_\rr$). \cite{BBGN19} use this to prove their amplification-by-shuffling lemma. The work by Feldman et al. \cite{FMT20} strengthens the result by, roughly speaking, performing the noise/data decomposition on an input-by-input basis.

\subsubsection{Other Protocols for Binary Sums}
We remark that there are shuffle protocols which have properties not present in $\rr$. These are achieved by leveraging the power of multiple messages. Table \ref{tab:binary-sums} presents their most salient features.

\begin{table}[h]
\renewcommand*{\arraystretch}{1.8}
    \centering
    \begin{tabular}{cccc}
        Error & No. Messages per User & Advantage over $\rr$ & Source \\ \hline
        $O\paren{\frac{1}{\eps^2} \log \frac{1}{\delta}}$ & 2 & If sum is 0, estimate is 0 with prob. 1 & \cite{BC19} \\ 
        $O\paren{\frac{1}{\eps} \sqrt{\log \frac{1}{\delta}} }$ & $O\paren{\frac{1}{\eps^2} \log \frac{1}{\delta}}$ $*$ & Symmetric noise & \cite{BCJM20} \\
        $O\paren{\frac{1}{\eps^{3/2}} \sqrt{\log \frac{1}{\eps}} }$  & $O\paren{\frac{1}{\eps} \log n}$ & $\delta=0$ & \cite{GGK+20} \\
        $O\paren{\frac{1}{\eps}}$ & $1+O\paren{\frac{1}{\eps^2 n}\log^2 \frac{1}{\delta}}$ $*$ & Optimal error & \cite{GKMP20} \\
        $O\paren{\frac{1}{\eps}}$ & $\tilde{O}\paren{\poly\paren{n,\frac{1}{\eps}}}$ & $\delta=0$ \emph{and} optimal error & \cite{CY21} \\ \hline
    \end{tabular}
    \caption{Shuffle protocols for binary sums. Each message is one bit. ``$*$'' denotes a bound that holds in expectation over the randomness of all users.}
    \label{tab:binary-sums}
\end{table}


We note that the protocol by Cheu \& Yan \cite{CY21} pays a large price in message complexity in order to achieve both optimal error and pure differential privacy. It is unclear if this price is necessary. Prior work by Ghazi, Golowich, Kumar, Manurangsi, Pagh, and Velingker\cite{GGK+20} and Ghazi, Kumar, Manurangsi, and Pagh \cite{GKMP20} were able to achieve only one of the two properties with much fewer messages.

\subsection{Histograms}
\label{sec:histogram}
In this setting, each user has one value in the set $[d]$. Let $c_j$ denote the count of $j$ in the input dataset. The objective is to privately compute a vector $(\tilde{c}_1, \dots, \tilde{c}_d)$ such that the $\ell_\infty$ distance from $(c_1,\dots,c_d)$ is small. In other words, the output's maximum error should be low. This error must grow with $d$ under local privacy:
\begin{thm}[Bassily \& Smith \cite{BS15}]
Let $\cP$ be an $(\eps,\delta)$-locally private protocol. If $\cP$ reports a histogram that has $\ell_\infty$ error $\alpha$ with constant probability, then $\alpha = \Omega(\frac{1}{\eps} \sqrt{n \log d})$
\end{thm}

In contrast, it is possible to have error independent of $d$ under robust shuffle privacy:
\begin{thm}[Balcer et al. \cite{BC19,BCJM20}]
\label{thm:histogram}
There is a shuffle protocol that satisfies $(\eps,\delta)$-robust differential privacy and outputs a histogram that has $\ell_\infty$ error $O(\frac{1}{\eps^2} \log \frac{1}{\delta})$ with constant probability.
\end{thm}
\begin{proof}
A simple way to obtain a private histogram is to perform $d$ different binary sums. A union bound suffices to upper bound the maximum magnitude of error. Basic composition ensures that we only pay a factor of 2 in the privacy parameters, since changing a user's value from $x$ to $x'$ only affects the counts of those two values.

In the shuffle model, the protocol executions can be done in parallel: each user simply labels their messages with the execution number. To be precise, let $\cR_j$ be a binary sum randomizer that counts the occurrences of $j$. If $\cR_j(x_i)$ outputs messages $a$ and $b$, user $i$ reports the tuples $( j,a )$ and $( j, b )$. To estimate $c_j$, we can feed the messages into the corresponding analyzer function $\cA_j$.

A side-effect of this reduction approach is that the union bound may create a dependence on $d$. For example, if we use $\rr$ for frequency estimation, the $\ell_\infty$ error after the union bound has a $\sqrt{\log d}$ term. But we can avoid this dependence by using $\zsum$, a binary sum protocol which guarantees noiseless estimation when the input is $(0,\dots,0)$. As such, the elements with nonzero frequency will be the only ones with noisy estimates. But there are only $\leq n$ of these, so the union bound is over $\leq n$ protocol executions instead of $d$.

We present the local randomizer of $\zsum$ below. $r$ is a parameter to be determined.
\[
\cR_\zsum(x_i) := (x_i, \Ber(r))
\]

\underline{Robust Privacy}: As with $\rr$, it suffices to prove privacy of the sum of the messages from honest users. But this quantity is exactly $\sum_{i=1}^{n/2} x_i + \eta$, where $\eta$ is drawn from the distribution $\Bin(n/2,r)$. And by Lemma \ref{lem:b_noise}, it suffices to choose $r= 1- \tfrac{\kappa}{\eps^2 n} \cdot \log \tfrac{1}{\delta}$ for $(O(\eps), \delta)$ privacy.\footnote{If $n < \tfrac{2\kappa}{\eps^2} \cdot \log \tfrac{1}{\delta}$, notice that $r>1/2$. In this case, honest users can simply opt to report (0,0). Perfect privacy is achieved at the price of error $n = O(\tfrac{1}{\eps^2} \cdot \log \tfrac{1}{\delta}) $}

\underline{Accuracy}: Now we define the analyzer $\cA_\zsum$.
\[
\cA_\zsum(\vec{y}) := \begin{cases}
0 & \textrm{if } \sum y_{i,1} + y_{i,2} \leq n \\
\sum y_{i,1} + y_{i,2}-nr & \textrm{otherwise}
\end{cases}
\]

First consider the case where $\sum x_i = 0$. Because $\eta \sim \Bin(n, r)$ has maximum value $n$, $\pr{}{\sum y_{i,1} + y_{i,2} \leq n}=1$ so there is zero error.

Now consider the case where $\sum x_i = 0$. We can use a Chernoff bound to argue that $|\eta - nr| = O(\sqrt{n(1-r)\log n})$ with probability $1/10n$. If we do not truncate, subtracting $nr$ removes bias so that error has magnitude $O(\sqrt{n(1-r)\log n}) = O(\frac{1}{\eps}\log \frac{1}{\delta})$. Otherwise, error is exactly $\sum x_i$. But truncation will not occur when $\sum x_i = \Omega(\frac{1}{\eps^2} \log \frac{1}{\delta})$: in this case, $\sum x_i + \eta > \sum x_i + n - O(\frac{1}{\eps^2} \log \frac{1}{\delta})$ so that $\sum y_{i,1} + y_{i,2} > n$.
\end{proof}

In Appendix \ref{apdx:comm}, we give a more technically involved protocol that has improved communication complexity. We also summarize alternative histogram protocols

\subsection{Uniformity testing}
In $\alpha$-\emph{uniformity testing}, we assume each user has an i.i.d. sample from some probability distribution $\bD$ over $[d]$. The objective is to report ``uniform'' with probability 2/3 when $\bD=\bU$ and ``not uniform'' with probability 2/3 when $\tv{\bD}{\bU} > \alpha$. The minimum number of users needed to ensure those two conditions hold is the \emph{sample complexity} of the protocol. Under local privacy, this must scale at least linearly with $d$.

\begin{thm}[Acharya et al. \cite{ACFT19}]
If an $\eps$-locally private protocol performs $\alpha$-uniformity testing, then its sample complexity is $\Omega(d/\alpha^2\eps^2)$.
\end{thm}

But under robust shuffle privacy, it has been shown that the sample complexity has a leading term of $d^{2/3}$ instead of $d$. Balcer, Cheu, Joseph, and Mao \cite{BCJM20} give the first protocol to achieve that bound. Canonne and Lyu \cite{CL22} streamline its analysis and describe a \emph{single-message} protocol of their own using privacy amplification. Both protocols demand approximate differential privacy.

The testing protocol by Cheu and Yan \cite{CY21} attains pure differential privacy while maintaining the same leading $d^{2/3}$ term. It follows much the same ``recipe'' as that of Balcer et al. \cite{BCJM20} which has two parts: a core testing protocol whose sample complexity scales with $d^{3/4}$ and a domain compression lemma that lets us reduce the sample complexity to one that scales with $d^{2/3}$. \cite{CY21}'s core tester uses a pure DP counting protocol while \cite{BCJM20} relies on an approx. DP counting protocol.

\begin{thm}[Cheu \& Yan \cite{CY21}]
There is a multi-message protocol that satisfies $\eps$-robust shuffle privacy and solves $\alpha$-uniformity testing with sample complexity\footnote{Special thanks to Cl{\'{e}}ment Canonne for simplifying the big-Oh expression.} $$ O\paren{ \frac{d^{3/4}}{\alpha\eps} + \frac{d^{1/2}}{\alpha^2}}.$$
\end{thm}

\begin{proof}
As previously mentioned, much of this construction is borrowed from Balcer et al. \cite{BCJM20}. We note that we will take $n\sim \Pois(m)$ and upper bound $m$. This ``Poissonization'' has the effect of making the random variables $c_1,\dots, c_d$ mutually independent, which simplifies the analysis.

Cai et al. \cite{CDK17} give a recipe for private uniformity testing under Poissonization. First, compute a private histogram $(\tilde{c}_1, \dots, \tilde{c}_d)$. Then, compute the test statistic
\begin{equation}
\label{eq:z-prime}
Z'(\tilde{c}_1, \dots, \tilde{c}_d) := \frac{d}{m} \sum_{j=1}^d (\tilde{c}_j - m/d)^2-\tilde{c}_j    
\end{equation}
The final step is to prove that this statistic is small when the data distribution is uniform but large when it is $\alpha$-far from uniform, which means we can distinguish the two cases with a threshold test.

Amin et al. \cite{AJM19} give the following procedure to analyze $Z'$. If we let $\eta_j$ be the noise in $\tilde{c}_j$ introduced by privacy, then we rewrite $Z'$ as
\begin{align*}
\eqref{eq:z-prime} &= \frac{d}{m} \sum_{j=1}^d (c_j + \eta_j - m/d)^2 - c_j - \eta_j\\
    &= \underbrace{\frac{d}{m} \sum_{j=1}^d (c_j - m/k)^2 - c_j}_{Z} + \underbrace{\frac{d}{m} \sum_{j=1}^d \eta^2_j}_{A} + \underbrace{\frac{2d}{m} \sum_{j=1}^d \eta_j \cdot (c_j-m/d)}_{B} - \underbrace{\frac{d}{m} \sum_{j=1}^d \eta_j}_{C}
\end{align*}
Analysis in Acharya et al. \cite{ADK15} imply bounds on term $Z$ in the two relevant cases: there is a constant $t$ and a function $f(\alpha,m)$ such that
\begin{enumerate}
\item when $\tv{\bD}{\bU}>\alpha$, $Z > t\cdot f(\alpha,m)$ with constant probability 
\item when $\bD=\bU$, $Z\leq f(\alpha,m)$ with constant probability
\end{enumerate}
If we prove the two statements below
\begin{enumerate}[label=(\roman*)]
    \item When $\tv{\bD}{\bU}>\alpha$, $A+B+C > 0$ with constant probability
    \item When $\bD=\bU$, $A+B+C< (t-1)\cdot f(\alpha,m)$ with constant probability
\end{enumerate}
then combining with 1. and 2. implies that the value $t\cdot f(\alpha,m)$ serves as a threshold that successfully separates the two cases with constant probability.

Balcer et al. \cite{BCJM20} describe a binary sum protocol which produces estimates with zero-mean symmetric noise.\footnote{At a high level, each user sends a random number of $\Ber(1/2)$ messages. The aggregate number of such bits is guaranteed to be $\Theta(\tfrac{1}{\eps^2}\log \tfrac{1}{\delta})$ with $1-O(\delta)$ probability.} A corollary is that there is a private histogram protocol where each $\eta_j$ is an independent sample from zero-mean symmetric noise. (i) follows from the fact that the probability of $\eta_j>0$ is 1/2. (ii) follows from Chebyshev's inequality and the moments of $\eta_j$.

Cheu \& Yan \cite{CY21} follow precisely the same template except they deploy a binary sum protocol that satisfies pure differential privacy ($\delta=0$) and $O(1/\eps)$ error.
\end{proof}

We now sketch how to reduce the sample complexity dependence on $d$ from $d^{3/4}$ to $d^{2/3}$. The technique is due to Acharya, Canonne, Han, Sun, and Tyagi \cite{ACHST20} and Amin et al. \cite{AJM19} (itself a generalization of a similar technique from Acharya et al. \cite{ACFT19}) The idea is to reduce the size of the data universe $[d]$ by grouping random elements and then performing the test on the smaller universe $[\hat{d}]$. The randomized grouping also reduces testing distance---partitions may group together elements with non-uniform mass to produce a group with near-uniform overall mass, thus hiding some of the original distance---but the reduction in universe size outweighs this side-effect.

\begin{lemma}[Domain Compression \cite{ACHST20, AJM19}]
\label{lem:partition}
Let $\bD$ be a distribution over $[d]$. For any partition $G$ of $[d]$ into $\hat{d} < d$ groups $G_1,\dots,G_{\hat{k}}$, let $\bD_G$ be the distribution over $[\hat{k}]$ with probability mass function $\pr{}{\bD_G = \hat{j}} := \sum_{j \in G_{\hat{j} }} \pr{}{\bD = j}$. If $G$ is chosen uniformly at random, then with probability $\geq 1/954$ over $G$,
$$\tv{\bD_G}{\bU} \geq \tv{\bD_G}{\bU} \cdot \frac{\sqrt{\hat{d}} }{477\sqrt{10d}}.$$
\end{lemma}

Public randomness can be used to create the partition $G$. Users can then replace their data $j$ with the partition $\hat{j}$ it belongs to. Running the protocol on the transformed dataset (with distance parameter $\hat \alpha := \alpha \tfrac{\sqrt{\hat d}}{477\sqrt{10d}}$) gives the final uniformity tester below:

\begin{theorem} [Cheu \& Yan \cite{CY21}]
Fix any $\eps = O(1)$, and $0 < \alpha < 1$. There exists a protocol that is $\eps$-robustly shuffle private and solves $\alpha$-uniformity testing with sample complexity
$$O\paren{\frac{d^{2/3}}{\alpha^{4/3} \eps^{2/3}} + \frac{d^{1/2}}{\alpha \eps} + \frac{d^{1/2}}{\alpha^2}}.$$
\end{theorem}

\subsection{Pointer-Chasing}
The \emph{pointer chasing problem} is denoted $\mathrm{PC}(k,\ell)$ where $k,\ell \in \N$. A problem instance is a set $\{(1,\vec{a}),(2,\vec{b})\}$, where $\vec{a}$ and $\vec{b}$ are permutations of $[\ell]$. A protocol \emph{solves $\mathrm{PC}(k,\ell)$ with sample complexity} $n$ if, given $n$ independent samples drawn uniformly with replacement from any problem instance $\{(1,\vec{a}),(2,\vec{b})\}$, it outputs the $k$-th integer in the sequence $a_1,b_{a_1},a_{b_{a_1}}\dots$ with constant probability.

Joseph, Mao, and Roth show that the sample complexity of $\mathrm{PC}(2,\ell)$ under local privacy must scale at least linearly with $\ell$.
\begin{thm}[Joseph et al. \cite{JMR19}]
\label{thm:pc-lower}
If an $(\eps,\delta)$-locally private protocol solves $\mathrm{PC}(2,\ell)$ with sample complexity $n$ then $n = \Omega(\ell)$.
\end{thm}

In stark contrast, the sample complexity under shuffle privacy is \emph{independent of} $\ell$:
\begin{thm}[Balcer \& Cheu \cite{BC19}]
There is a $8\cdot(\ell!)^2$-message protocol that satisfies $(\eps, \delta)$-robust shuffle privacy and solves $\mathrm{PC}(2,\ell)$ with sample complexity $O(\frac{1}{\eps^2}\log \frac{1}{\delta})$.
\end{thm}
\begin{proof}
Let $\pi(\ell)$ denote all permutations of $[\ell]$. Observe that the tuples $(1,\vec{a}),(2,\vec{b})$ are elements of the universe $\{1,2\} \times \pi(\ell)$ which has size $2\cdot \ell!$. We can solve the problem once we have a protocol that singles out $(1,\vec{a})$ and $(2,\vec{b})$ from the universe with constant probability.

Balcer \& Cheu argue that the task of privately identifying $(1,\vec{a})$ and $(2,\vec{b})$ with constant probability is $O(\frac{1}{\eps^2}\log \frac{1}{\delta})$. By a straightforward concentration argument, it suffices to have $O(t)$ samples to ensure $(1,\vec{a})$ and $(2,\vec{b})$ each appear $\geq t+1$ times with constant probability. Taking universe size $d = 4\cdot (\ell !)^2$, we then use the histogram protocol built atop $\zsum$ (Theorem \ref{thm:histogram}). When $t = \Omega( \frac{1}{\eps^2}\log \frac{1}{\delta} )$, it will report nonzero frequencies for $(1,\vec{a})$ and $(2,\vec{b})$ but zero for every other element in the universe.
\end{proof}





\section{Separations between Central \& Shuffle Privacy}
\label{sec:lower}
There are known separations between the (one-round) shuffle model and the central model. The proofs thus far require some natural structural constraint.

\subsection{Single-message Shuffle Privacy}
\label{sec:lower-single}
The first class of lower bounds hold for protocols wherein each user sends exactly one message with probability 1.\footnote{The lower bounds also hold in the case where users send \emph{at most} one message. This is proven by a simple transformation: send a dummy symbol $\bot$ to denote the no-message event.} We begin with a negative result for bounded-value sums proved by Balle, Bell, Gasc{\'{o}}n, and Nissim \cite{BBGN19}.

\begin{thm}[Balle et al.  \cite{BBGN19}]
If a single-message shuffle protocol satisfies $(\eps,\delta)$ differential privacy for $n$ users and computes bounded-value sums, then the mean-squared error must be $\Omega(n^{1/3})$.
\end{thm}
In contrast, the centrally private Laplace mechanism achieves mean-squared error of $O(1/\eps^2)$.

\medskip

The techniques used to prove the above are specific to bounded-value sums. A more general technique is to study what happens when we remove the shuffler from a single-message protocol. This takes us to what we can call \emph{removal lemmas}
\begin{lem}[Balcer \& Cheu \cite{BC19}]
\label{lem:removal-bc}
If a single-message protocol $\cP=(\cR,\cA)$ satisfies pure shuffle privacy, then removing the shuffler leaves behind a pure locally private protocol. Specifically, $\cR$ must satisfy $\eps$-differential privacy on its own whenever the shuffle protocol as a whole is $\eps$-private.
\end{lem}
This means that under pure differential privacy, the single-message shuffle model is \emph{exactly equivalent} to the local model. So all separations between the central and local models hold here as well.

But it is clear from $\rr$ that this exact equivalence does not hold for approximate shuffle privacy. The following removal lemma accommodates the relaxation.
\begin{lem}[Cheu et al. \cite{CSU+19}]
\label{lem:removal-csuzz}
If a single-message protocol $\cP=(\cR,\cA)$ satisfies $(\eps,\delta)$-shuffle privacy for $n$ users, then $\cR$ must satisfy $(\eps+\ln n, \delta)$-differential privacy on its own.
\end{lem}
Thus, we can invoke any local model lower bound that holds for $(\eps+\ln n, \delta)$ privacy. As an example, the recipe implies the following lower bound on the error of histograms.
\begin{thm}[Ghazi et al. \cite{GGK+19}]
Any single-message protocol that satisfies $(1,o(1/n))$-shuffle privacy and estimates a $d$-bin histogram with $\ell_\infty$ error $n/10$ must have $n = \Omega( \frac{\log d}{\log \log d})$.
\end{thm}
In contrast, there is a central model algorithm where $n=O(1)$ suffices for the same privacy and accuracy regimes.

\subsection{$m$-message Shuffle Privacy}
\label{sec:lower-multi}

A natural idea is to somehow extend the removal lemma from the single-message case to the $m$-message case. But there are differentially private shuffle protocols whose randomizers are \emph{not} differentially private. For example, an adversary can recover the input of $\cR_\zsum$ by simply looking at the first bit of the output. Other examples can be found in Appendix \ref{apdx:brittle}.

Despite this hurdle, two works manage to prove lower bounds for $m$-message protocols. These lower bounds make the simplifying assumption that the local randomizer sorts (or shuffles) its output messages before giving them to the shuffler. This does not affect accuracy or privacy because the local sorting (or local shuffling) is undone by the shuffler anyway.

\subsubsection{Approach 1}

One paper by Beimel, Haitner, Nissim, and Stemmer \cite{BHNS20} obtains a bound on the mutual information between the output of an $m$-message randomizer and uniformly random input.

\begin{lem}
Let $\cP=(\cR,\cA)$\footnote{The original statement allows for different users to run different randomizers, but we omit that degree of freedom for simplicity} be an $m$-message $(\eps,\delta)$-shuffle private protocol and let $Z_1,\dots,Z_n \in \cX$ be (possibly correlated) random variables. In the execution of $\cP$ on input $Z_1,\dots,Z_n$, let $Y_i$ be the (sorted) output of the $i$-th user and let $W$ denote the public randomness. For any $i\in[n]$, if $Z_i$ is uniformly random over $\cX$, then $$I(Y_i,W; Z_i) = O\left((en)^m \cdot \left(\eps^2 + \frac{\delta}{\eps}\log |\cX| + \frac{\delta}{\eps} \log \frac{\eps}{\delta} \right) + m \log n\right).$$
\end{lem}
\begin{proof}[Proof Sketch]
Given $(\eps,\delta)$-shuffle private protocol $\cP=(\cR,\cA)$, we can create a $(\eps,\delta)$-locally private randomizer $\cR_{\cP}$: on input $x, W$, obtain $nm$ messages by executing $(\cS\circ\cR^n)(U_1, U_2, \dots, U_{n-1}, x)$ where $U_i$ is uniformly random, and then output a random (sorted) subset of $m$ messages. Privacy follows from post-processing.

Now let $Y'_i \gets \cR_\cP(Z_i,W)$. Prior work has shown that $I(Y'_i,W;Z_i)=O(\eps^2 + \tfrac{\delta}{\eps}\log |\cX| + \tfrac{\delta}{\eps} \log \tfrac{\eps}{\delta})$. Then we use the fact that $Y'_i$ coincides with $Y_i$ with probability $\binom{nm}{m}^{-1}$.
\end{proof}

The above lemma is then used to obtain a lower bound for the \emph{common element} problem. Refer to \cite{BHNS20} for the full details.

\subsubsection{Approach 2}

A paper by Chen, Ghazi, Kumar, and Manurangsi \cite{CGKM21} takes a different approach. They define a relaxation of differentially private algorithms---called \emph{dominated algorithms}--- and then argue that the local randomizer of a shuffle private protocol satisfies that definition.

\begin{defn}[Chen et al. \cite{CGKM21}]
An algorithm $\cR:\cX\times \zo^* \to \cY$ is $(\eps,\delta)$-dominated if there exists a distribution $\bD$ such that for all $x\in \cX$, all $w\in \zo^r$, and all $Y\in\cY$, $\pr{}{\cR(x,w) \in Y} \leq e^\eps \cdot \pr{}{\bD \in Y} + \delta$
\end{defn}
Notice that the above definition is a one-sided variant of differential privacy since it does not require the probability mass function of $\cR(x,w)$ to dominate that of $\bD$.

\begin{lem}
If $\cP=(\cR,\cA)$ is an $m$-message $(\eps,\delta)$-shuffle private protocol, then $\cR$ is $(\eps + m\ln(en), \delta)$-dominated.
\end{lem}

Using the above, Chen et al. derive a lower bound for parity learning:

\begin{thm}
If $\cP$ is a $m$-message shuffle protocol that solves $d$-dimensional parity learning, then its sample complexity is $\Omega(2^{d/(m+1)})$.
\end{thm}

In contrast, Kasiviswanathan, Lee, Nissim, Raskhodnikova, and Smith \cite{KLNRS08} show that centrally private parity learning is possible with just $O(d)$ samples.
 
\subsection{Robust Shuffle Privacy}
\label{sec:lower-robust}
The third class of lower bound applies to robustly shuffle private protocols. To obtain these results, we again develop reductions, but this time to the online model. Briefly, an online algorithm receives user data one at a time and updates its internal state upon reading each input. The algorithm produces output when the stream ends.

How do we define privacy in the online model? Dwork, Naor, Pitassi, Rothblum, and Yekhanin \cite{DNPRY10} propose \emph{pan-privacy}: for any time $t$, the joint distribution of the internal state at time $t$ and the output should be differentially private. This models one-time violations of the algorithm's integrity (i.e. a hack, a subpoena, or a change in ownership).

Balcer, Cheu, Joseph, and Mao \cite{BCJM20} describe a generic transformation from robust shuffle privacy to pan-privacy that preserves accuracy for many statistical problems. Thus, existing lower bounds that hold under pan-privacy---for the distinct elements and uniformity testing problems---carry over to robust shuffle privacy. Cheu \& Ullman \cite{CU21} and Nissim \& Yan \cite{NY21} obtain new lower bounds for pan-private selection and parity learning, which again imply lower bounds for robust shuffle privacy. This second batch of results imply exponentially large separations in sample complexity between robust shuffle privacy and central privacy. Refer to Table \ref{tab:lower} for an overview of these results.

In the thesis by Cheu \cite{C21}, the recipe is somewhat simplified. The key observation is that lower bounds for pan-privacy typically only require the privacy of the internal state and not that of the state-output pair. \cite{C21} uses \emph{internal privacy} to refer to this weaker notion. Transforming robustly shuffle private protocols to internally private algorithms is a little easier than transforming them to pan-private algorithms, while still producing the same results.\footnote{It also avoids a technical limitation of the original transformation, which is that privacy parameter needs to be known when a third of users participate.}

In the following lemma, $\bU$ is any distribution\footnote{As the symbol suggests, it is typically the uniform distribution} over the data universe $\cX$ and let $\bU^n$ be the corresponding product distribution over $\cX^n$. For any other distribution $\bD$, let $\bD_{(p)}$ be the mixture $p\cdot \bD + (1-p)\cdot \bU$.

\begin{lem}[Balcer et al. \cite{BCJM20}, Cheu \cite{C21}]
\label{lem:reduction}
Let $\cP=(\cR,\cA)$ be an $(\eps,\delta)$-robustly shuffle private protocol. There is an $(\eps,\delta)$-internally private algorithm $\cQ_\cP$ such that
\begin{equation}
\label{eq:equivalence}
    \dtv(\cQ_\cP(\bU^{n/2}), \cP(\bU^n))=0    
\end{equation}
and, for any distribution $\bD$ over $\cX$,
\begin{equation}
\label{eq:dilution}
    \dtv(\cQ_\cP(\bD^{n/2}), \cP(\bD^n_{(1/4)}) ) < 1/6.
\end{equation}
\end{lem}
\begin{proof}[Proof Sketch]
The online algorithm's initial internal state will be the output of $(\cS \circ \cR^{n/2})$ run on $n/2$ i.i.d. samples from $\bU$. Each time a user's data point is read, the algorithm will execute $\cR$ on it and add the messages to the internal state (inserted in some random position). This ensures internal privacy because any internal state is equivalent to the output of the shuffler when the protocol is run on (at least) $n/2$ data points.

The output of $\cQ_\cP$ is simply the execution of $\cA$ on the final state. \eqref{eq:equivalence} is immediate from the construction. To obtain \eqref{eq:dilution}, we begin with the observation that the final internal state consists of messages produced by running the protocol on independent samples from $\bU,\dots,\bU,\bD,\dots,\bD$. This looks almost like $\bD_{(1/2)}$ except that the number of samples from $\bD$ should be binomial. We correct this by slightly modifying the transformation: replace the first $\Bin(n/2,q)$ user data with samples from $\bU$. $q$ is chosen so that the shuffled set of samples approximates samples from $\bD_{(1/4)}$. The modification does not invalidate our preceding arguments.
\end{proof}

\begin{table}
\centering
\caption{Comparison of impossibility results for robust shuffle privacy with centrally private algorithms. $d$ and $\alpha$ are dimension and error parameters, respectively. $k$ is the number of inputs to the learned parity function. For simplicity, we use $\eps = \hat{\eps}(1/2)$ and $\delta =\hat{\delta}(1/2)$. $*$ indicates that $\delta\log (\binom{d}{\leq k}/\delta) \ll \alpha^2\eps^2 / \binom{d}{\leq k}$.}
\label{tab:lower}

\renewcommand{\arraystretch}{1}
\begin{tabular}{cccc}
  & & Robust Shuffle Privacy & Central Privacy\\ \hline
 \rule{0ex}{3.5ex}\vspace{1ex} \multirow{2}{*}{Additive Error of} & \multirow{2}{*}{Distinct Elements} &  $\Omega\left( \sqrt{\frac{d}{\eps}} + \frac{1}{\eps} \right)$ & $O\left(\frac{1}{\eps}\right)$ \\
 \vspace{.5ex} & & \cite{BCJM20} ($n\geq 2d$) & \cite{DMNS06} \\ \hline
 \rule{0ex}{3.5ex}\vspace{1ex} & \multirow{2}{*}{Uniformity Testing} & $\Omega \left( \frac{d^{2/3}}{\alpha^{4/3} \eps^{2/3}}+ \frac{\sqrt{d}}{\alpha^2} + \frac{1}{\alpha \eps} \right)$ & $O\left(\frac{\sqrt{d}}{\alpha^2} + \frac{\sqrt{d}}{\alpha \eps} + \frac{d^{1/3}}{\alpha^{4/3} \eps^{2/3}} + \frac{1}{\alpha \eps} \right)$ \\
 \vspace{.5ex} Sample & & \cite{BCJM20} ($\delta=0$) & \cite{ASZ18} \\ \cline{2-4}
 \rule{0ex}{3.5ex}\vspace{1ex} Complexity of & \multirow{2}{*}{Parity Learning} & $\Omega\left( \sqrt{\binom{d}{\leq k}} / \alpha \eps  \right)$ & $O(\log \binom{d}{\leq k})$  \\
 \vspace{.5ex}& & \cite{CU21} agnostic, \cite{NY21} realizable $*$ & \cite{KLNRS08} \\  \hline
 \end{tabular}
\end{table}

\section{The Promise of Interactivity}
Thus far, we have limited our attention to one-round shuffle protocols. We shall now explore what shuffle protocols can do with multiple rounds of communication and how they stack up against centrally private algorithms.

\subsection{Sequential Interactivity (S.I.)}

To start, it will help to understand sequentially interactive local protocols. Here, each user sends only one message but the randomizer of user $i$ can depend on the \emph{transcript} generated by users $1,\dots,i-1$. This is useful when implementing private iterative methods like gradient descent. Strong separations are known to exist between one-round and sequentially interactive local privacy. Joseph, Mao, and Roth \cite{JMR19} show that two rounds suffice to solve pointer-chasing $\mathrm{PC}(2,\ell)$ with sample complexity $O_\eps(\log \ell)$. This is exponentially smaller than the lower bound of $\Omega_\eps(\ell)$ in the one-round case (Theorem \ref{thm:pc-lower}).

Given that S.I. provably enhances the local model, how can we adapt it to the shuffle model?

\paragraph{Approach 1.} One option is to re-interpret the shuffler as an anonymity service: users are shuffled u.a.r. and then the analyzer deploys a sequentially interactive local protocol.\footnote{An equivalent interpretation is that, at the beginning of each round of the S.I. local protocol, a middle-man samples a random user without replacement.} The recent amplification-by-shuffling lemma by Feldman, McMillan, and Talwar \cite{FMT20} holds in this version of the model. Notice that if the randomizer does not get updated over time, we are just running a one-round single-message shuffle protocol. Also observe that it is not possible to run multi-message protocols in this variant of the shuffle model, since the random permutation is limited to the users and not the messages.

\paragraph{Approach 2.} An alternative way to adapt S.I. is to simply run one-round shuffle protocols on disjoint batches of users. The $i$-th protocol can depend on the transcript from protocols $1,\dots,i-1$ and can be multi-message. Summarized in Table \ref{tab:si}, two recent works have described protocols in this model. Tenenbaum, Kaplan, Mansour, Stemmer \cite{TKMS21} study the multi-arm bandit problem. The authors give cumulative regret bounds that match those of the central model up to logarithmic factors. Cheu, Joseph, Mao, and Peng \cite{CJMP21} focus instead on the problem of stochastic convex optimization (SCO). They describe a one-round vector summation protocol that is repeatedly called inside gradient descent algorithms.

\begin{table}[]
    \centering
    \caption{Comparison of positive results in the S.I. shuffle model with central model counterparts. For brevity, we suppress the term $\sum_{a\in[k]:\Delta_a>0} \frac{\log T}{\Delta_a}$ present in both MAB bounds and the term $1/\sqrt{n}$ in the SCO bounds. SCO bounds also omit logarithmic factors, as well as convexity and smoothness parameters.}
    \label{tab:si}
    
    \renewcommand{\arraystretch}{1}
    \begin{tabular}{cccc}
      & & S.I. Shuffle Privacy & Central Privacy\\ \hline
     \rule{0ex}{3.5ex}\vspace{1ex} \multirow{2}{*}{Regret of} & $k$-arm & $O\left( \frac{k}{\eps}\sqrt{\log \tfrac{1}{\delta} } \log T \right)$ & $O(k/\eps)$ \\ 
     \vspace{.5ex}  &  bandit & \cite{TKMS21} & \cite{TD16} \\ \hline
     \rule{0ex}{3.5ex}\vspace{1ex}  & Convex, Non-Smooth & $O(d^{1/3}/\eps^{2/3}n^{2/3})$ &  \\ 
     \vspace{.5ex} SCO & Convex, Smooth  & $O(d^{2/5}/\eps^{4/5}n^{4/5})$ & $O(\sqrt{d}/\eps n)$ \\
     \vspace{.5ex} error & Strongly Convex, Non-Smooth  &  $O(d^{2/3}/\eps^{4/3}n^{4/3})$ & \cite{BST14} \\
     \vspace{.5ex}  & Strongly Convex, Smooth  & $O(d/\eps^2 n^2)$ & \\ \hline
    \end{tabular}
\end{table}

\subsection{Full Interactivity (F.I.)}

In fully interactive local protocols, a user can communicate with the analyzer multiple times. The transcript of all the user's messages must be differentially private.

We can adapt F.I. to the shuffle model in the following way: run one-round shuffle protocols on batches of users that are not necessarily disjoint. The transcript of a fully interactive shuffle protocol is the entire list of the outputs of the shuffler. As with local protocols, this transcript must be differentially private. As an example, Cheu et al. \cite{CJMP21} give a SCO protocol that relies on this ability to query a user multiple times.

Beimel et al. \cite{BHNS20} describe a very powerful transformation that shows fully interactive shuffle private protocols can be as powerful as centrally private ones(!)
\begin{thm}
Let $\cM$ be an arbitrary (central model) randomized algorithm. Assuming an honest majority and semi-honest corruptions, there exists a two-round fully interactive shuffle protocol $\cP_\cM$ that simulates $\cM$.
\end{thm}
\begin{proof}[Proof Sketch]
The idea is to simulate an information-theoretically secure multi-party computation protocol by Applebaum, Brakersky, and Tsabary (ABT), the source of the honest majority requirement. The MPC protocol relies on secure channels of communication; to simulate these channels in the shuffle model, Beimel et al. use one-time pads.

We begin with a simple building block: Alice and Bob want to agree on one random bit, with one party designated as ``leader.'' As usual, the adversary's view is limited to the output of the shuffler. Suppose both Alice and Bob each flip one fair coin and send their bits. By examining the output of the shuffler, each party can learn what the leader sampled.\footnote{We can use the analyzer as a referee to relay the shuffler's output. Alternatively, we could model the shuffler as having the ability to broadcast its output (as done by Beimel et al.).} However, if both have 0 or both have 1, the adversary learns both their bits. This has a $1/2$ chance of occurring, so they repeat the process enough times to drive the probability down. Note that these repetitions can be done in parallel by labeling each bit with a repetition number. When there are $n>2$ users, we label each message with the pair of users who will read them.

Naively combining the above key agreement with ABT leads to a three-round protocol (one for key agreement and two for ABT). Beimel et al. show how to use the leftover hash lemma to send a message and the pad at the same time, reducing the number of rounds to two.
\end{proof}


\section{Open Questions}

\paragraph{How can we close the gap between the amplification and removal lemmas?} The best-known amplification lemma has constraints on the number of users and privacy parameters. In particular, they do not apply to $(\ln en, \delta)$-private randomizers. Randomizers with these parameters are created by the removal lemma by Cheu et al (Lemma \ref{lem:removal-csuzz}). If the removal lemma guarantees could be tightened (or the amplification constraints could be loosened), then we would be able to show that a single-message shuffle protocol is intrinsically robust: when only a constant fraction $\gamma$ of users participate, amplification of the randomizer's privacy guarantee would give us a concrete privacy parameter for $\cS\circ\cR^{\gamma n}$.

\paragraph{What is the optimal sample complexity of uniformity testing under approximate differential privacy?} We have matching upper and lower bounds for testing under pure differential privacy. Prior work has shown that, in the central model, pure d.p. is not a stronger constraint on the sample complexity of a binary decision problem than approximate d.p. But is this the case for pan-privacy? Robust shuffle privacy?

\paragraph{What are the limits of S.I. protocols?} It appears difficult to perform the same level of simulation as done in the fully interactive setting. There may be a way to adapt the strong lower bounds developed by Joseph et al. \cite{JMR19,JMNR19}. Note that we can ask this question for both approaches of defining S.I. shuffle protocols.


\bibliographystyle{plain}
\bibliography{references}

\appendix
\section{Other Models of Distributed Differential Privacy}
\label{apdx:other-models}
Here, we discuss other distributed models and compare them with the shuffle model.

\paragraph{Secure Aggregation Model.} In this model, there is a trusted service or functionality called the aggregator. Much like the shuffle model, users send their messages to the aggregator who then reports a value to the analyzer. The aggregator's output is the sum of user messages, modulo some modulus. \footnote{In the language of Cheu and Yan \cite{CY21}, the shuffler and the aggregator are two realizations of a \emph{secure intermediary}.}

Ishai, Kushilevitz, Ostrovsky, and Sahai \cite{IKOS06} show that we can perform secure aggregation the shuffle model: each user samples a set of $m$ values uniformly over sets that add up to their sensitive value. The protocol guarantees that, for large enough $m$ and any inputs $\vec{x},\vec{x}\,'$ that have the same sum, the distribution of the multiset of all user shares does not change significantly between $\vec{x},\vec{x}\,'$. We will not provide the security proof for space, but we will use the construction in Appendix \ref{apdx:comm}.

Conversely, it is not difficult to show that a protocol in the shuffle model implies one in the secure aggregation model. A user could encode the set of $m$ messages they intend to send as an integer where the $j$-th digit is the count of $j$ in that message set. Adding up these encodings will yield the histogram of all user messages, which contains exactly the same information as the shuffled set of them.

\paragraph{Multi-central Model.} In the work by Steinke \cite{Ste20}, we find a model where a user can communicate with any subset of $k>1$ servers but they are only guaranteed at least one honest server. The honest server(s) must communicate in a manner such that the view of the dishonest servers is insensitive to any single user's contribution.

Steinke shows that secret sharing can be combined with server-side noise to compute differentially private sums. One limitation of this protocol is that a malicious user can influence the sum by a magnitude as large as the modulus of secret sharing, which needs to be at least as large as $n$. Recently, Talwar \cite{Tal22} offers an alternative protocol which verifies the magnitude of the user's contribution before adding it.

Multi-central protocols are at least as powerful as shuffle protocols with finite communication complexity. This follows from (1) the earlier observation that we can losslessly transform a multi-set of messages into a (large) integer and (2) secret sharing for summation.

Instead of that transformation-based approach, Steinke argues that we can \emph{directly execute} shuffle protocols using public key cryptography (onion encryption). Furthermore, differentially private selection can be performed using $\log d$ samples when given access to an MPC implementation of argmax. In contrast, the techniques used in \cite{CU21} imply selection demands $d$ samples under robust shuffle privacy.

\paragraph{Two-Party Protocols.} The work by McGregor, Mironov, Pitassi, Reingold, Talwar, and Vadhan \cite{MMPRTV10} formalizes the following scenario: there are two servers, two disjoint datasets (no user appears in both), and server $j\in\zo$ has exclusive access to dataset $j$. The servers communicate with one another across rounds. An honest server $j$ should interact with (potentially malicious) server $1-j$ such that the transcript is simulatable by a differentially private algorithm on dataset $j$.

This model is very close to the central model, since any ``i.i.d.-style'' problem like mean estimation, uniformity testing, and learning can simply be solved by each server on their own. Indeed, it is not hard to see that protocols in the distributed models we have seen so far can be simulated by two-party protocols. Still, McGregor et al. derive a lower bound on the \emph{inner product} problem, where the goal is to estimate the inner product between the two datasets. $O(1/\eps)$ error is possible in the central model via the Laplace mechansim but $\Omega(\sqrt{n})$ is necessary for the two-party model (and the others that can be simulated by it)
\section{Message Complexity and Communication Complexity}
\label{apdx:comm}

In any real implementation of a shuffle protocol, users will have to transmit their messages across a network. The two critical metrics are the number of messages sent by each user and the total number of bits they consume. We use \emph{message complexity} to refer to the former and \emph{communication complexity} to refer to the latter. In this Appendix, we give an overview of protocols that are designed to minimize one or both of these quantities. We also take a glimpse at lower bounds.

\subsection{Communication-efficient Bounded-value Sums}
In this setting, users have values in the interval $[0,1]$ and the objective is to privately compute their sum. It is possible to use a binary sum protocol for this problem: a fixed-point representation can transform a continuous value into a set of zeroes and ones, upon which we perform the local ranomization. A longer fixed-point representation reduces the rounding error, but increases the noise needed for privacy; two works \cite{CSU+19,CJMP21} show that $\sqrt{n}$-long representation suffices.

The downside of the above approach is that the message complexity ---and thus the communication complexity--- scales with $\sqrt{n}$. To rectify this, Balle et al. \cite{BBGN19-2,BBGN19-3} and Ghazi et al. \cite{GPV19,GMPV19} use a different reduction that leads to a logarithmic communication complexity.

\begin{thm}
There is an $(\eps,\delta)$-shuffle private protocol for bounded-value sums with error $O(\frac{1}{\eps})$ where each user sends $1+O(\frac{\log(1/\delta)}{\log(n)})$ messages, each consisting of $O(\log n)$ bits.
\end{thm}
\begin{proof}[Proof Sketch]
At a high level, the goal is to simulate the symmetric geometric distribution $\SG(\eps)$, also known as the discrete Laplace distribution. In the central model, adding such noise suffices for pure differential privacy.

The first step is to equate a sample from the $\SG(\eps)$ with the sum of $n$ samples from another distribution $\bD_\eps$. This property is called \emph{infinite divisibility}, most obvious in the Gaussian distribution. The next step is to recall the modular arithmetic (or secure aggregation) protocol $\cP_\mmod = (\cR_\mmod, \cA_\mmod)$ by Ishai et al. \cite{IKOS06}. It ensures that two input datasets with the same sum (modulo some modulus) cause the protocol to produce a shuffled set of messages that are $\delta$-close in statistical distance. Finally, we define $\cR$ to be the execution of $\cR_\mmod$ on $y_i \gets x_i+\eta$ where $\eta\sim \bD_\eps$.

If an adversary can only recover $\sum y_i$, then we will have $\eps$-differential privacy. Due to our use of $\mmod$, the output of the shuffler $(\cS\circ\cR^n)(x_1,\dots,x_n)$ is $\delta$-close to the output of the algorithm $(\cS\circ\cR^n_\mmod)(\sum y_i ,0,\dots,0)$. This closeness suffices for approximate differential privacy. The error bound $O(\frac{1}{\eps})$ because we are simulating the geometric mechanism and sums exceed the modulus with very low probability (assuming the modulus is large).

Refer to Balle et al. \cite{BBGN19-3} and Ghazi et al. \cite{GMPV19} for analyses of the message complexity of $\mmod$.
\end{proof}

As an aside, Cheu and Yan \cite{CY21} follow much the same template, except that the security property of their $\mmod$ replaces statistical distance with one derived from the definition of pure differential privacy. But this variant protocol demands exponentially more bits.

\subsection{Almost-communication-efficient Histograms with Domain-Independent Error}

Here, we describe a protocol that has the same asymptotic error as the protocol by Balcer \& Cheu but reduced communication complexity.

\begin{thm}
\label{thm:eff-hist}
Fix any $T\in \N$ and privacy parameters $0<\eps,\delta = O(1)$. There exists an $(\eps,\delta)$-private shuffle protocol which estimates histograms up to $\ell_\infty$ error $O(T^2\log(T/\delta)/\eps^2)$ with at least $99/100-\delta$ probability and consumes $O\paren{\frac{T^3d^{1/T}}{\eps^2}\log \frac{T}{\delta}}$ messages of length $O\paren{\log Tn + \frac{1}{T}\log d}$.
\end{thm}

The construction proceeds in two steps. We first make an inefficient but accurate protocol, then describe a technique to reduce its communication complexity.

\subsubsection{An opt-in protocol}

Derived from conversations with Maxim Zhilyaev, this protocol reports private histograms such that the $\ell_\infty$ error is $O( \log(1/\delta)/\eps^2 )$ with $1-\delta$ probability. At a high level, each user ``opts-in'' to contributing noise to the count of each universe element. Much like in the analysis of shuffled randomized response, the size of this opt-in set only depends on the privacy parameters (and not $n$ or $d$). This will in fact determine the maximum error of the histogram.

\medskip

The first message user $i$ sends is their true value $x_i$. Their second message is a bit $b_i$ drawn from $\Ber(p)$ where $p = \Theta( \log(1/\delta)/\eps^2 n )$. This bit determines whether or not the user opts-in: if $b_i=1$, they will also flip $d$ fair coins. If the $j$-th coin is heads, then they send $j$ as yet another message.

To prove this protocol is differentially private, let $H$ be the set of all users $i$ where $b_i=1$. We leverage the following concentration result: for sufficiently large $n$, $|H| \geq \frac{\kappa}{\eps^2}\log \frac{1}{\delta}$ with probability $\geq 1- \delta$ where $\kappa$ is the constant from Lemma \ref{lem:b_noise}. This implies differential privacy: the noise in the frequency of each $j$ is an independent sample from $\Bin(|H|,1/2)$ and $|H|$ is sufficiently large to ensure that additive noise offers $(\eps,\delta)$-privacy.

Error is also low. Because the noise on any bin is drawn from $\Bin(|H|,1/2)$, we have that the $\ell_\infty$ error is $\leq |H| = O( \log(1/\delta)/\eps^2 )$ with probability $\geq 1-\delta$. Note that the analyzer can compute $|H|$ by simply adding up the one-bit messages. Also, we can derive the ``zero-maps-to-zero'' property from Balcer-Cheu by truncating estimates to 0 if they are at most $|H|$. 

\medskip

What is the communication complexity? Each user sends a one-bit message alongside at least one $\log_2 d$-bit message. The number of $\log_2 d$-bit messages sent by a user is a random variable with expectation $1+p\cdot (d/2) = 1+\Theta(d\log(1/\delta)/\eps^2 n)$. Contrast this with $1+\Theta(d\cdot (1-\log(1/\delta)/\eps^2 n))$ in the protocol by Balcer \& Cheu.

\subsubsection{Count-Min Template}
This meta-protocol samples random hash functions and repeatedly executes a subroutine for computing histograms on hashed data. The number of repetitions determines both the privacy parameters and the size of the hashed domain. See pseudocode in Algorithms \ref{alg:cm-randomizer} and \ref{alg:cm-analyzer}

\begin{algorithm}

\caption{$\cR_{CM}$ a local randomizer for histograms}
\label{alg:cm-randomizer}

\KwIn{$x\in[d]$; parameters $T,\hat{d}\in\N$; randomizer $\cR:[\hat{d}]\to \cY^*$}
\KwOut{$\vec{y}\in ([T]\times \cY)^*$}

Obtain hash functions $\{h^{(t)} : [d]\to [\hat{d}]\}$ from public randomness.

Initialize $\vec{y}\gets \emptyset$

\For{$t\in[T]$}{
	Compute $\vec{y}^{(t)} \gets \cR(h^{(t)}(x))$

	\For{$y \in \vec{y}^{(t)}$} {
		Append $(t,y)$ to $\vec{y}$	
	}
}

\Return{$\vec{y}$}
\end{algorithm}

\begin{algorithm}

\caption{$\cA_{CM}$ an analyzer for histograms}
\label{alg:cm-analyzer}

\KwIn{$\vec{y}\in ([T]\times \cY)^*$; parameters $T,\hat{d}\in\N$; analyzer $\cA:\cY^*\to \R^{\hat{d}}$}
\KwOut{$\vec{z}\in\R^d$}

Obtain hash functions $\{h^{(t)} : [d]\to [\hat{d}]\}$ from public randomness.

\For{$j\in[d]$}{
	$z_j\gets \infty$
}

\For{$t\in[T]$}{
	Initialize $\vec{y}^{(t)}\gets \emptyset$
	
	\For{$(t',y) \in \vec{y}$} {
		Append $y$ to $\vec{y}^{(t)}$ if $t'=t$
	}
	
	Compute $\hat{z}^{(t)} \gets \cA(\vec{y}^{(t)})$
	
	\For{$j\in[d]$}{
		$\hat{j} \gets h^{(t)}(j)$
	
		$z_j\gets \min(z_j,\hat{z}^{(t)}_{\hat{j}})$
	}
}

\Return{$\vec{z}$}
\end{algorithm}

\begin{thm}
Fix any number of users $n$, domain size $d$ and natural number $T$. Let $\cP=(\cR,\cA)$ be any shuffle protocol for computing $d$-bin histograms where (1) each user sends, in expectation, $M(d)$ messages of length $L(d)$ (2) $(\eps,\delta)$-privacy is offered to any user and (3) with probability $\geq 1-\beta$, the $\ell_\infty$ error is $\alpha(d,\beta)$. If we instantiate $\cP_{CM}=(\cR_{CM},\cA_{CM})$ with that $\cP$ and parameters $T,\hat{d} \gets \lceil n\cdot (100d)^{1/T} \rceil$, then
\begin{enumerate}
\item
    each user sends, in expectation, $T\cdot M(\hat{d})$ messages of length $L(\hat{d})+\log T$
\item
	$\cP_{CM}$ is $\left(T\eps, T\delta \right)$-private
\item
	with probability $\geq 99/100-\beta$, the $\ell_\infty$ error is $\alpha(\hat{d},\beta/T)$.
\end{enumerate}
\end{thm}

\begin{proof}
Since the randomizer executes $\cR$ exactly $T$ times on hashed user data, labeling messages with the execution number each time, Item 1 is immediate from substitution and Item 2 follows from basic composition.

To prove Item 3, let $E_j$ denote the event that there is a hash function $h^{(t)}$ such that a user's value $j$ experiences no collisions with another user: formally, $\exists t ~ \forall j'\in \vec{x},j'\neq j~ h^{(t)}(j)\neq h^{(t)}(j')$. When this event occurs, observe that the count of $h^{(t)}(j)$ in the hashed dataset is precisely the count of $j$ in the original dataset. Otherwise, the count of $h^{(t)}(j)$ is at least as large as $j$. Given that the analyzer $\cA$ reports estimates with max error $\alpha(\hat{d},\beta)$ with probability $\geq 1-\beta$, a union bound implies the minimum over all $T$ repetitions can only be wrong by $\alpha(\hat{d},\beta/T)$ with probability $\geq 1-\beta$. Thus, it suffices to bound the probability that $E_j$ does not occur for some $j$.
\begin{align*}
\pr{\vec{h}}{\neg E_j} &= \pr{\vec{h}}{\forall t ~ \exists j' \in \vec{x}~ h^{(t)}(j)= h^{(t)}(j')} \\
	&= \pr{\vec{h}}{\exists j' \in \vec{x}~ h^{(t)}(j)= h^{(t)}(j')}^T \\
	&\leq (n\cdot \pr{\vec{h}}{h^{(t)}(j)= h^{(t)}(j')})^T \\
	&= (n/\hat{d})^T = (1/(100d)^{1/T})^T = 1/100d \\
\therefore \pr{\vec{h}}{\exists j ~ \neg E_j} &\leq 1/100 \qedhere
\end{align*}
\end{proof}

\subsection{Communication-efficient Histograms \& Range queries}
In Table \ref{tab:histograms}, we compare the above protocol with the protocols by Ghazi et al \cite{GGK+19}. The communication complexities of those protocols have only a logarithmic dependence on $n,d$. They combine compression techniques that found success in the local model with the privacy blanket notion.

The table presents two other histogram protocols. The first uses the parallel-counts template that we used in Section \ref{sec:histogram}, but now with the binary sum protocol presented by Ghazi, Kumar, Manurangsi, and Pagh \cite{GKMP20}. The expected message complexity of this protocol vanishes with $n$, so that a large userbase counteracts a large dimension $d$. The second also has a vanishing message complexity, but with a faster rate. Each user in this protocol by Cheu and Zhilyaev \cite{CZ21} randomizes the one-hot encoding of their data, as well as a small number of $(0,\dots,0)$ strings. These fake users contribute just enough cover noise to protect real users.

\begin{table}[h]
\renewcommand*{\arraystretch}{1.8}
\caption{Shuffle protocols for histograms. All take $\delta>0$. We assume $\delta < 1/\log d$ for results from \cite{GGK+19}. $T$ is a natural number. The notation $\tilde{O}(\dots)$ suppresses nested logarithms. \vspace{1ex}}
    \label{tab:histograms}
    \centering
    \begin{tabular}{cccc}
    Source & Error & Messages per User & Bits per Message\\ \hline
     Thm. \ref{thm:eff-hist} & $O\!\left(\frac{T^2}{\eps^2} \log \frac{T}{\delta} \right)$ &
     $O\paren{\frac{T^3d^{1/T}}{\eps^2}\log \frac{T}{\delta}}$ & $O\paren{\log Tn + \frac{1}{T}\log d}$ \vspace{2.2pt} \\ \hline
    \multirow{2}{*}{\cite{GGK+19}} & $\tilde{O}\!\left(\frac{1}{\eps} \sqrt{\log^3 d  \log \frac{1}{\delta} }\,\right)$ & $\tilde{O}\!\left(\frac{1}{\eps^2}\log^3 d  \log \frac{1}{\delta} \right)$ & $O(\log n + \log \log d)$ \\
    & $O\!\left(\log d + \frac{1}{\eps} \sqrt{\log d \log \frac{1}{\eps \delta} \,}\right)$ &
     $O\!\left(\frac{1}{\eps^2}\log\frac{1}{\eps\delta}\right)$ &
     $O(\log n \log d)$ \vspace{2.2pt} \\ \hline
     \cite{GKMP20} & $O(\frac{1}{\eps}\log d)$ & $1+O(\frac{d}{\eps^2n} \log^2 \frac{1}{\delta})$ & $O(\log d)$ \\
     \cite{CZ21} & $O(\log d + \frac{1}{\eps} \sqrt{\log d  \log \frac{1}{\delta}} ) $ & $1+O(\frac{\log d}{n} + \frac{1}{\eps^2 n}\log \frac{1}{\delta})$ & $d$ \\
     \end{tabular}
\end{table}

In \cite{GGK+19}, Ghazi et al. also explain how to use their protocols in a black-box way to solve the range-query problem. In this setting, data is drawn from $[k]^d$ and the objective is to estimate the number of points in a given rectangle. Refer to Table \ref{tab:range-counting} for a summary of the results.

\begin{table}[h]
\renewcommand*{\arraystretch}{1.8}
    \caption{Shuffle protocols for range queries. All take $\delta>0$. $n\leq k^d$ for neatness}
    \label{tab:range-counting}
    \centering
    \begin{tabular}{cccc}
         Technique & Error& Messages per User & Bits per Message \vspace{2.2pt}\\ \hline
         Count-Min & $O(\frac{1}{\eps}\log^{2d+3/2} (k^d) \log \frac{1}{\delta})$ & $O(\frac{1}{\eps^2}\log^{3d+3}(k^d) \log \frac{1}{\delta})$ & $O(\log n +\log(d\log k))$ \\
         Hadamard & $O(\frac{1}{\eps}\log^{2d+1/2} (k^d) \log \frac{1}{\eps\delta})$ & $O(\frac{1}{\eps^2}\log^{2d}(k^d) \log \frac{1}{\eps \delta})$ & $O(\log (n) \cdot d\log k)$\vspace{2.2pt} \\ \hline
    \end{tabular}
\end{table}

\subsection{A Lower Bound for Binary Sums}
A result by Ghazi et al. \cite{GGK+20} states that every communication-bounded shuffle protocol must imply some local protocol with a nontrivial privacy guarantee:
\begin{lem}[Ghazi et al. \cite{GGK+20}]
Suppose $\cP=(\cR,\cA)$ satisfies $(O(1),0)$-shuffle privacy and each user sends $m$ messages of $\ell$ bits. Then the local randomizer $(\cS\circ \cR^1)$ satisfies $(0,1-2^{-O(m^2 \ell)})$-differential privacy.
\end{lem}
By way of the local model, this implies a lower bound for binary sums:
\begin{coro}[Ghazi et al. \cite{GGK+20}]
If an $m$-message shuffle protocol satisfies $O(1)$-differential privacy and computes binary sums up to error $o(\sqrt{n})$, then $m^2 \ell = \Omega(\log n)$.
\end{coro}

\subsection{A Lower Bound for Vector Sums}

In the context of the secure aggregation model, Chan, Choquette-Choo, Kairouz, and Suresh \cite{CCKS22} prove the following lemma regarding finite-precision representations of vectors:
\begin{lem}
Fix any algorithm $M$ that takes as input a $d$-dimensional unit vector (in Euclidean space) and outputs $b$ bits. If there exists an algorithm $A$ where $\ex{}{\norm{(A\circ M)(x)-x}^2_2} \leq \alpha$, then $b \geq \frac{d}{2}\log_2 (1/\alpha)$. If we also have that $\ex{}{(A\circ M)(x)-x}=0$, then $b=\Omega(d/\alpha)$.
\end{lem}

We can use the above to obtain a lower bound on the communication complexity of any secure intermediary protocol for vector sums 

\begin{coro}
Let $P=(R,I,A)$ be any secure intermediary protocol (e.g. $I$ is a shuffler or aggregator). If $P$ privately estimates vector means with near-optimal $\ell_2$ error---formally, $\ex{}{\norm{P(\vec{x})-\frac{1}{n}\sum x_i}^2_2} = \tilde{O}(d/n^2\eps^2)$---then $R$ must be supported on a set of size at least $2^b$ for $b = \tilde{\Omega}(\max(d\log (n^2\eps^2/d),1))$. If the estimate is unbiased, then $b=\tilde{\Omega}(\min(d,n^2\eps^2))$.
\end{coro}

The shuffle protocol by Cheu, Joseph, Mao, and Peng \cite{CJMP21} is unbiased and has near-optimal error. It executes $d$ scalar mean protocols in parallel, each one consuming $O(\sqrt{n} + \frac{1}{\eps^2}\log \frac{1}{\delta})$ messages of $O(\log d)$ bits. In the regime where $d < n^2\eps^2$, the communication complexity is suboptimal.

To improve that bound, the authors suggest replacing the scalar mean subroutine with that of Balle Bell Gasc\'{o}n and Nissim. This alternative protocol is biased and, once we re-scale parameters and label messages for use in the vector mean protocol, consumes $O\paren{\paren{\log\frac{d}{\delta} + \log(n+\frac{1}{\eps}\log d)}/\log n}$ messages of $\log (d (n+\frac{1}{\eps}\log d))$ bits. So this modification is within polylogarithmic factors of the general lower bound

\section{Shuffle Protocols with Brittle Privacy}
\label{apdx:brittle}
Here, we describe two protocols which satisfy non-trivial shuffle privacy but are not robust to a single drop-out.

\begin{thm}
There exists a protocol $\cP=(\cR,\cA)$ such that $(\cS\circ\cR^n)$ satisfies pure differential privacy but $(\cS\circ\cR^{n-1})$ \emph{does not} satisfy pure differential privacy.
\end{thm}
\begin{proof}
Define $\cR:\zo\to \{1\}^*$ such that the length of the output (number of messages) is uniformly random over $\{0,\dots, n+2\}$ on input 0 and uniformly random over $\{0,1,n+1,n+2\}$ on input 1.

We first show that $(\cS\circ\cR^n)$ is $\eps$-differentially private for a finite value of $\eps$. This is achieved by arguing that, for every input $\vec{x}$, the length of $(\cS\circ\cR^n)(\vec{x})$ has support $G = \{0,\dots,n^2+2n\}$. We use the notation $\mathrm{supp}(|(\cS\circ\cR^n)(\vec{x})|) = G$. This equivalence holds if and only if the two following statements are true: (i) the length of $(\cS\circ\cR^n)(\vec{x})$ must be some member of the set $G := \{0,\dots,n^2+2n\}$ and (ii) each integer in $G$ has a nonzero probability of being the length.

(i) is immediate from the specification of $\cR$: the length is maximized when all users send $n+2$ messages and minimized when they send no messages. To prove (ii), we perform case analysis over $\vec{x}$.



When $\vec{x}=0^n$, we shall use induction over the elements of $G$ in order. The base case is immediate: $\pr{}{|(\cS\circ\cR^n)(0^n)|=0} = \pr{}{|\cR(0)|=0}^n > 0$. For the inductive step, we are given that $\pr{}{|(\cS\circ\cR^n)(0^n)|=g}>0$ for some $g\in G - \{n^2 + 2n\}$ and we show that $\pr{}{|(\cS\circ\cR^n)(0^n)|=g+1}>0$. There must be a vector $\vec{g} \in \{0,\dots,n+2\}^n$ such that $\sum g_i = g$ and $\prod_{j=1}^n \pr{}{|\cR(0)| = g_j} > 0$. Because $g < n^2 + 2n$, there must be some index $i$ such that $g_i < n+2$. Hence, define $\vec{g}\,'$ such that $g'_i = g_i + 1$ and $g'_j = g_j$ for all $j\neq i$. Now we have that $\pr{}{|(\cS\circ\cR^n)(0^n)|=g+1} \geq \prod_{j=1}^n \pr{}{|\cR(0)| = g'_j} > 0$.

When $\vec{x}=1^n$, the proof is similar except the inductive step proceeds via case analysis. If $0 \in \vec{g}$, we simply create $\vec{g}\,'$ by changing the 0 to 1. If $n+1 \in \vec{g}$ we create $\vec{g}\,'$ by changing the $n+1$ to $n+2$. Otherwise, there is some integer $k\geq 0$ such that $\vec{g}$ consists of $k$ copies of $(n+2)$ and $n-k$ copies of 1. In this case, we construct $\vec{g}\,'$ which has $n-k-1$ copies of 0 and $k+1$ copies of $n+1$. In all cases, $\sum g'_j = 1 + \sum g_j$ and $\prod \pr{}{\cR(1)=g'_j} > 0$.

For any other choice of $\vec{x}$, the fact that $\mathrm{supp}(|\cR(1)|) \subset \mathrm{supp}(|\cR(0)|)$ implies $$\mathrm{supp}(|(\cS\circ\cR^n)(1^n)|)\subseteq \mathrm{supp}(|(\cS\circ\cR^n)(\vec{x})|) \subseteq \mathrm{supp}(|(\cS\circ\cR^n)(0^n)|)$$ so that all the supports are precisely $G$.

Now we show that $(\cS\circ\cR^{n-1})$ cannot satisfy pure differential privacy. Consider the neighboring inputs $\vec{x} := 0^{n-1}$ and $\vec{x}\,' := 0^{n-2}1$. There is a non-zero probability that $(\cS\circ\cR^{n-1})(\vec{x})$ has length $n$. However, this is impossible when the input is $\vec{x}\,'$, so the likelihood ratio is unbounded.
\end{proof}

\begin{thm}
There exists a protocol $\cP=(\cR,\cA)$ such that $(\cS\circ \cR^n)$ satisfies approximate differential privacy, but $(\cS\circ\cR^{n-1})$ does not satisfy \emph{any} differential privacy.
\end{thm}
\begin{proof}
Define $\cR:\zo\to \{1\}^*$ such that the length of the output is uniform over $\{0,1\}$ on input 0 and uniform over $\{n,n+1\}$ on input 1.

We first show that $(\cS\circ\cR^n)$ is $(\eps,\delta)$-differentially private for a finite value of $\eps$ and $\delta<1$. This is achieved by arguing that, for any neighboring $\vec{x}\sim\vec{x}\,'$, the support of $(\cS\circ\cR^n)(\vec{x})$ intersects with that of $(\cS\circ\cR^n)(\vec{x}\,')$. Let $k$ be the number of times 0 occurs in $\vec{x}$; without loss of generality, assume that the number of times 0 occurs in $\vec{x}\,'$ is $k+1$. We have that 
\begin{align*}
&\pr{}{|(\cS\circ\cR^n)(\vec{x})| = n^2-kn}\\
\geq{}& \pr{}{|(\cS\circ\cR^k)(0^k)| = 0} \cdot \pr{}{|(\cS\circ\cR^{n-k})(1^{n-k})| = (n-k)\cdot n}\\
>{}& 0
\end{align*}
and that
\begin{align*}
&\pr{}{|(\cS\circ\cR^n)(\vec{x}\,')| = n^2-kn}\\
\geq{}& \pr{}{|(\cS\circ\cR^k)(0^k)| = k} \cdot \pr{}{|\cR(0)|=1} \cdot \pr{}{|(\cS\circ\cR^{n-k-1})(1^{n-k-1})| = (n-k-1)\cdot (n+1)}\\
>{}& 0
\end{align*}

Now we argue that $(\cS\circ\cR^{n-1})$ cannot satisfy any degree of differential privacy. Given $\vec{x}=0^{n-1}$ and $\vec{x}\,'=0^{n-2},1$, the maximum length of $(\cS\circ\cR^{n-1})(\vec{x})$ is $n-1$ while the minimum length of $(\cS\circ\cR^{n-1})(\vec{x}\,')$ is $n$. Hence, we have neighboring inputs but the supports of the induced distributions are disjoint.
\end{proof}

\end{document}